# TEMPORAL DYNAMIC CONVOLUTIONAL NEURAL NETWORK FOR TEXT-INDEPENDENT SPEAKER VERIFICATION AND PHONEMIC ANALYSIS

*Seong-Hu Kim, Hyeonuk Nam, Yong-Hwa Park*

Department of Mechanical Engineering, Korea Advanced Institute of Science and Technology, Korea

## ABSTRACT

In the field of text-independent speaker recognition, dynamic models that adapt along the time axis have been proposed to consider the phoneme-varying characteristics of speech. However, a detailed analysis of how dynamic models work depending on phonemes is insufficient. In this paper, we propose temporal dynamic CNN (TDY-CNN) that considers temporal variation of phonemes by applying kernels optimally adapting to each time bin. These kernels adapt to time bins by applying weighted sum of trained basis kernels. Then, an analysis of how adaptive kernels work on different phonemes in various layers is carried out. TDY-ResNet-38(×0.5) using six basis kernels improved an equal error rate (EER), the speaker verification performance, by 17.3% compared to the baseline model ResNet-38(×0.5). In addition, we showed that adaptive kernels depend on phoneme groups and are more phoneme-specific at early layers. The temporal dynamic model adapts itself to phonemes without explicitly given phoneme information during training, and results show the necessity to consider phoneme variation within utterances for more accurate and robust text-independent speaker verification.

## 1. INTRODUCTION

Phonemes and acoustic characteristics of utterances change with the spoken text, and text-independent speaker recognition should recognize speakers regardless of the text. However, previous studies have shown that speaker embeddings and performance depend on phonemes [1-3]. This means that the phoneme configuration within the text affects the performance of text-independent speaker recognition. Thus, a model that extracts consistent speaker information regardless of phonemes is required for more accurate and robust text-independent recognition. From this insight, the model using an adaptive convolutional neural network (ACNN) in which the kernels adaptively change along time segments was proposed, and accurate speaker information was extracted regardless of phonemes [4]. It showed the applicability of dynamic models (or content-adaptive models) that adapt to temporal data to consider phoneme information in text-independent speaker recognition. However, kernels adaptive to time segments might not be enough to fully capture the information of different phonemes because not only the classes of phonemes but also the time length of them depends on the speaker's speaking style or speed. Therefore, we would like to propose a model that extracts speaker information by adapting to time bins, which are the smallest unit of spectrogram.

The dynamic models are suitable for improving text-independent speaker recognition performance in this context. There are few studies on implementing dynamic models on text-independent speaker recognition [4, 5] when compared to dynamic models applied in computer vision [6-12] and NLP [13, 14] applications. In image classification tasks, the characteristics of the adaptive kernels on various images were analyzed to examine how the performance improved [11]. It was observed that images with similar visual characteristics adopt the same basis kernels to form adaptive kernels. This result verified that dynamic models were trained to extract visual information according to the visual context of images. However, among the works applying dynamic models in speaker recognition tasks, no one has performed a detailed analysis of adaptive kernels in relation to phonemes.

In this paper, we propose a temporal dynamic convolutional neural network (TDY-CNN) for text-independent speaker verification and analyze how adaptive kernels change with phonemes. The main contributions of this work are as follows:

1. We propose CNN kernels adaptive to each time bin in order to effectively capture the time-varying information in utterances.

2. This is the first work to perform phonemic analysis of the temporal dynamic model for text-independent speaker verification.

3. We verified that adaptive kernels change with the acoustic characteristics of phonemes and extract speaker information regardless of phonemes while static kernels do not.

The remainder of the paper is organized as follows. Section 2 introduces the temporal dynamic convolutional neural network for text-independent speaker verification. Section 3 describes experiment setup and details. Section 4 shows the experiment results and analysis. Finally, Section 5 presents conclusions.

## 2. TEMPORAL DYNAMIC CONVOLUTIONAL NEURAL NETWORK

Content-adaptation of kernels can be achieved in two ways: directly generating kernels and adjusting trained kernels [15]. The former methods guarantee diversity of adaptive kernels when compared to the latter methods, but it is difficult to define criteria to analyze the relationship between adaptive kernels and phonemes

This work was supported by "Human Resources Program in Energy Technology" of the Korea Institute of Energy Technology Evaluation and Planning (KETEP), granted financial resource from the Ministry of Trade, Industry & Energy, Republic of Korea. (No. 20204030200050) composing the utterance. In the latter methods, adaptive kernels are obtained by weighted sum of trained basis kernels, and we could compare the weights of the basis kernels with respect to phonemes composing the utterance. In addition, the computational cost for applying attention weights on the basis kernels is much lower than the cost for directly generating kernels. Thus, based on the latter methods, we propose a temporal dynamic convolutional neural network (TDY-CNN) to apply an adaptive kernel for each time bin, which is the smallest unit of spectrogram. A basic idea of TDY-CNN is originated from the dynamic convolutional neural network (DY-CNN) [12] which ensemble the kernels with softmax constraint of attention weights.

DY-CNN uses content-adaptive kernels obtained by weighted sum of basis kernels using attention weights. Similarly, TDY-CNN efficiently applies adaptive convolution depending on time bins by changing the computation order as follows:

$$y_k(f,t) = W_k * x(f,t) + b_k \quad (1)$$

$$y(f,t) = \sigma\left(\sum_{k=1}^{K} \pi_k(t) \cdot y_k(f,t)\right) \quad (2)$$

where $x$ and $y$ are input and output of TDY-CNN module which depends on frequency feature $f$ and time feature $t$ in time-frequency domain data. As shown in Equation 1, $k$-th basis kernel $W_k$ is convoluted with $x$ and $k$-th bias $b_k$ is added. $y_k$ are aggregated using the attention weights $\pi_k(t)$ which depends on time bins. $K$ is the number of basis kernels, and $\sigma$ is an activation function ReLU. The attention weight $\pi_k(t)$ has a value between 0 and 1, and the sum of all basis kernels on a single time bin is 1 as the weights are processed by softmax.

Attention weights are calculated by applying two fully-connected layers followed by input flattened along the channel and frequency dimensions to fully consider phonemic information in it. We set the hidden layer size of the attention weight decision model to be 1/8 of the previous layer size. Softmax constraint compresses the space of aggregated kernels to train the attention weights with high accuracy using less kernels per layer [12]. In the end, TDY-CNN is equivalent to convoluting adaptive kernels on each time bin. By replacing conventional CNN with TDY-CNN, we could not only improve speaker verification performance but also analyze the attention weights of adaptive kernels with respect to phoneme which is the smallest unit of utterances along time.

## 3. EXPERIMENT SETUP AND DETAILS

### 3.1. Input representations

The 64-dimensional log Mel-spectrograms are extracted using a hamming window of width 25ms with step 10ms and number of fast Fourier transform 512. We randomly extract the Mel-spectrogram segment for about 2 seconds from each utterance and use it for training. Mean and variance normalization is performed on every frequency bin of the Mel-spectrogram.

### 3.2. Baseline architecture and loss

VGG-M and ResNet have been proposed in the field of image classification [16, 17] and also applied to the field of speaker recognition [18-22]. We use VGG-M and ResNet-34 as baseline models in the experiments. The channels of ResNet-34 are modified to half (ResNet-34(×0.5)) and a quarter (ResNet-34(×0.25)) to reduce computational cost. We aggregated frame-level embeddings using average pooling rather than self-attention pooling in order to apply uniform attention and backward gradients for all frame-level features. TDY-CNN module replaces all convolution modules except for the first convolution layer that extracts global features.

The models are trained using a loss function combining the Angular Prototypical loss with the vanilla softmax loss. This combination demonstrates better verification performances than using other loss functions or other combinations [21].

### 3.3. Implementation details and datasets

Our implementation[1] is based on PyTorch [23] and the training process was run on NVIDIA TITAN RTX. The Adam optimizer with a weight decay of $5\times10^{-5}$ is used. Initial learning rate was $10^{-3}$, and the learning rate was decreased by a factor of 0.75 every 15 epochs. Batch normalization was used with a fixed batch size of 800 and no data augmentation was used. For TDY-CNN, we reduced the temperature in softmax constraint from 31 to 1 linearly in the first 10 epochs [12]. The models are trained on 5994 speakers of Voxceleb2 development set [19], and we validate them using Voxceleb1 test set [18].

### 3.4. Evaluation metrics

We sampled ten 4-second-long segments with the same intervals from each test segment and computed the $10 \times 10 = 100$ cosine similarities between every pair of segments. Mean of 100 similarities is used as the final pairwise score [19, 21]. Based on the average similarity score, we calculate equal error rate (EER) and the minimum value of cost function $C_{det}$ as evaluation metrics of verification. EER is the rate at which both acceptance and rejection errors are equal. $C_{det}$ is a weighted sum of false-reject and false-accept error probabilities with parameters $C_{miss} = 1$, $C_{fa} = 1$ and $P_{target} = 0.05$ [18, 24].

## 4. RESULTS AND ANALYSIS

### 4.1. Text-independent speaker verification of TDY-CNNs

#### 4.1.1. The number of basis convolution kernels

The number of basis kernels $K$ is directly related to the model complexity and the speaker verification performance. Table 1 shows the speaker verification performance with different $K$. The performance improves as $K$ increases until $K = 6$, where the model yields the best performance with EER of 1.58%. However, the error rate was increased when $K = 8$. The reasons for performance decreasing are the difficulty of optimization for larger models and overfitting induced as the increase in representation power of models, just as in previous works [11, 12]. Thus, we chose $K = 6$ and continued analysis.

#### 4.1.2. Comparison of static model and utterance/frame-level dynamic model

It is necessary to verify whether temporal dynamic models are suitable for speaker verification when compared to dynamic models and static models. Table 2 shows the speaker verification performance of ResNet-34(×0.25) using static CNN, DY-CNN, and TDY-CNN with six basis kernels. DY-CNN and TDY-CNN outperform static CNN, and TDY-CNN performs the best. This demonstrates that dynamic models are more suitable for speaker verification than static models, and TDY-CNN, which considers frame-level speaker information, is the best for speaker verification, unlike DY-CNN, which considers only utterance-level speaker information.

---
[1] https://github.com/shkim816/temporal_dynamic_cnn/

**Table 1.** Speaker verification performance of temporal dynamic models according to number of basis kernels.

| TDY-ResNet-34(×0.25) | EER (%) | MinDCF |
|---|---|---|
| $K = 2$ | 1.99 | 0.140 |
| $K = 4$ | 1.62 | 0.128 |
| $K = 6$ | **1.58** | **0.116** |
| $K = 8$ | 1.69 | 0.133 |

**Table 2.** Speaker verification performances of models using dynamic convolution with frame-level and utterance level.

| Network | EER (%) | MinDCF |
|---|---|---|
| ResNet-34(×0.25) | 2.43 | 0.184 |
| DY-ResNet-34(×0.25) | 2.07 | 0.157 |
| TDY-ResNet-34(×0.25) | **1.58** | **0.116** |

**Table 3.** Speaker verification performances of the networks without data augmentation.

| Network | #Parm | EER (%) | MinDCF |
|---|---|---|---|
| VGG-M | 4.16M | 3.77 | 0.287 |
| TDY-VGG-M | 71.2M | 3.04 | 0.237 |
| ResNet-34(×0.25) | 1.86M | 2.43 | 0.184 |
| TDY-ResNet-34(×0.25) | 13.3M | 1.58 | 0.116 |
| ResNet-34(×0.5) | 6.37M | 1.79 | 0.134 |
| TDY-ResNet-34(×0.5) | 51.9M | **1.48** | **0.118** |
| ResNet-50 [19] | 67.0M | 3.95 | 0.429 |
| Thin ResNet-34 [20] | 12.4M | 2.87 | 0.310 |
| H/ASP [22] | 8.00M | 1.29 | 0.091 |
| ECAPA-TDNN [25] | 14.7M | **1.18** | **0.088** |

*4.1.3. Text-independent speaker verification results*

The text-independent speaker verification performance is compared between baseline models and temporal dynamic models with six basis kernels as shown in Table 3. TDY-ResNet-34(×0.5) has an EER of 1.48%, which is the best performance without data augmentation on this paper. All models to which TDY-CNN was applied show better performance than the models using static CNN. This shows that TDY-CNN improves the performance of general text-independent speaker verification models.

We also compared performance between the temporal dynamic models and previous studies [19, 20, 22, 25] trained with Voxceleb2 without data augmentation and tested with Voxceleb1. The proposed models show good verification performance but lag slightly behind the state-of-the-art performance. Also, the models using TDY-CNN have more parameters than other networks. The reason for such large model parameters is that we set TDY-CNN to obtain attention weights from frequency and channel information without pooling to consider phoneme information as much as possible. It might cause performance degradation with overfitting.

**4.2. Analysis of adaptive kernels and frame-level embeddings in relation to phonemes**

TDY-CNN changes kernels depending on the time bins and speakers without explicitly using phonemic information during training. Thus, we attempted to verify that the temporal dynamic model could effectively extract speaker information that is consistent over different phonemes by comparing the difference of adaptive kernels on phonemes. Since adaptive kernels are obtained by the weighted summation of basis kernels, we directly compared attention weights $\pi_k$ of basis kernels.

We analyzed the correlation between attention weights and phonemes in several layers using TIMIT dataset [26] which provides phoneme labels. The analysis is performed on the utterances by the same speaker in order to exclude the effect of different speakers on the adaptive kernel. The attention weights are extracted using trained TDY-ResNet-34(×0.25) with $K = 6$ on two speakers, FDAS1 with the largest variance of attention weights and MHRM0 with the smallest variance of attention weights. They are visualized for six phoneme classes (vowels, semivowels and glides, nasals, fricatives and affricates, stops, and closures) at three different layers (Layer 5, Layer 15, and Layer 30) using principal component analysis (PCA) applied on the attention weights on 6 basis kernels, as shown in Figure 1.

At layer 5, the distribution of attention weights can be divided into three groups: voiced sounds, stops, and fricative-likes. The voiced sound group consists of vowels, semivowels, glides, and nasals. Their pronounced sounds are dominated by the voice produced by vocal cords. However, unlike the voiced sound group, the fricative-like group consisting of fricatives and affricates is dominated by the turbulent sound caused by constriction in the mouth cavity. The stop group consisting of stops and closures involves abrupt and impulsive sounds that share no similarity with characteristics of voiced sound and fricative-like groups. The distributions of voiced sounds, fricative-likes, and stops are close within each group while the groups are distinctively distributed. Different mechanisms of pronunciation and acoustic characteristics cause these three groups to have different distributions of attention weights. We also observed that phonetic information is more actively used at the early layer. The boundary between the distribution of groups starts to seem vague at layer 15, and the distributions completely merge and groups become indistinguishable at layer 30. The overall results show that TDY-CNN chooses attention weights of basis kernels depending on specific groups. Since each group is distinguished by its acoustic features of phonemes, it clearly supports the assumption that kernels of TDY-CNN adapt to phonemes. In addition, frame-level speaker features have not only speaker information but also phoneme information, so attention weights are more phoneme-specific at earlier layers. On the other hand, the distribution of attention weights is not correlated with phonemes at later layers because only speaker information has been extracted through the layers for consistent speaker embeddings regardless of phonemes.

Moreover, it is necessary to check the correlation between speaker embeddings and phonemes. We extracted frame-level speaker embeddings for the previous two speakers by removing the average pooling layer of TDY-ResNet-34(×0.25) and baseline model ResNet-34(×0.25). The speaker embeddings are projected using t-SNE algorithm [27] and displayed in Figure 2. In the results using the baseline model, the speaker embeddings are well gathered, but the embeddings of nasals and fricatives within speaker MHRM0 are far from the center of the embedding group.

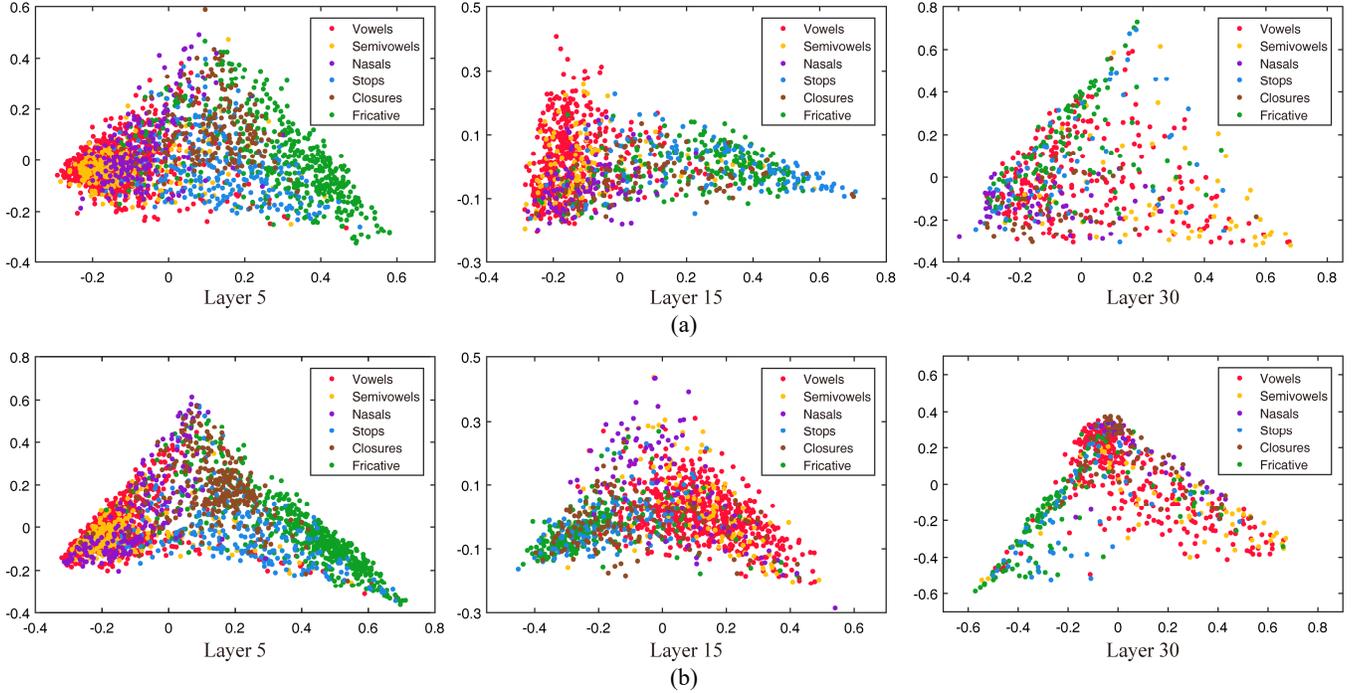

**Fig. 1.** Low-dimensional PCA projection of attention weights in speaker (a) MHRM0 and (b) FDAS1 for several layers and phoneme classes using TDY-ResNet-34(×0.25). Semivowels label contains glides, and fricative label contains affricative. The attention weights are more phoneme-specific at the early layer.

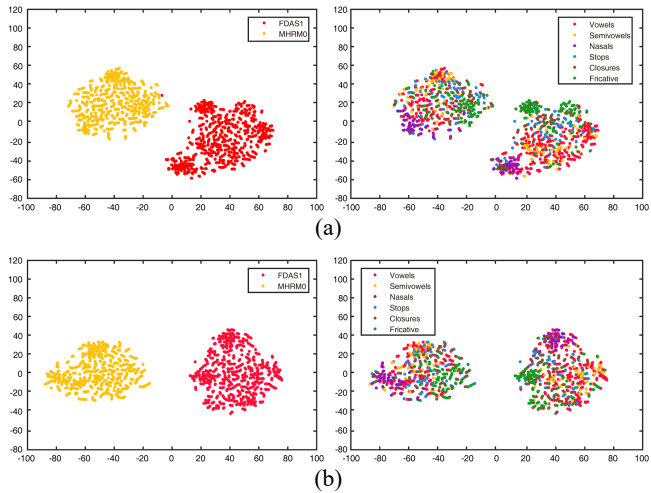

**Fig. 2.** Low-dimensional t-SNE projection of frame-level speaker embeddings of MHRM0 and FDAS1 using (a) baseline model ResNet-34(×0.25) and (b) TDY-ResNet-34(×0.25). Left column represents embeddings for different speakers, and right column represents embeddings for different phoneme classes.

On the contrary, embeddings by TDY-ResNet-34(×0.25) are closely gathered regardless of phoneme groups. This shows that the temporal dynamic model extracts consistent speaker information regardless of phonemes. However, since phoneme information has not been completely excluded, it can be seen that embeddings of phonemes are still grouped together. Therefore, we confirmed that TDY-CNN, which shows better speaker verification performance, adapts to phonemes and extract consistent speaker embeddings regardless of phonemes.

## 5. CONCLUSION

In this paper, we propose TDY-CNN to extract consistent speaker information on different time bins for text-independent speaker verification. Adaptive kernels of TDY-CNN are obtained by weighted sum of basis kernels with softmax constraint on attention weights. The model using TDY-CNNs with six basis kernels achieved better verification performance than the static models did, but it requires more model parameters when compared to static CNN. Also, we verified that adaptive kernels change depending on the groups of phonemes with similar acoustic characteristics without explicitly given phoneme information. The phoneme-specific characteristics of kernels are prominent at the early layers where the phonetic features still remain. At the later layer, only speaker information remains with little phoneme information, so the adaptive kernels of later layers are not correlated with acoustic characteristics of phonemes. Thus, models with TDY-CNN effectively extract consistent speaker information regardless of phonemes using phoneme-adaptive kernels and improve speaker verification performance. This work is the first to analyze how temporal dynamic models work depending on time bins and phonemes. The results indicate that temporal dynamic models are suitable and consideration of phoneme information is crucial in text-independent speaker verification.

In future works, we plan to analyze the structure of adaptive kernels depending on phonemes and propose optimal dynamic network with less parameters for text-independent speaker verification. In addition, we intend to analyze performance and relationship between kernels and phonemes when training the temporal dynamic model using not only speaker information but also phoneme information.